\RequirePackage{fix-cm}
\documentclass[letterpaper]{article}

\usepackage{spconf,amsmath,amssymb,graphicx,booktabs,url}
\usepackage[T1]{fontenc}
\usepackage{microtype}
\usepackage{balance}
\usepackage{tikz}

\usetikzlibrary{arrows.meta,positioning,calc,fit,backgrounds}
\usepackage[hidelinks]{hyperref}

\hypersetup{
    pdftitle={Beyond the Stability-Plasticity Frontier in Streaming Target Speaker Extraction},
    pdfauthor={},
    pdfsubject={ICASSP 2027}
}

\newcommand{\st}{\mathbf{s}_t}
\newcommand{\et}{\mathbf{e}_t}
\newcommand{\W}{\mathbf{W}}
\newcommand{\anchor}{\mathbf{a}}
\newcommand{\norm}{\operatorname{norm}}

\newcommand{\pointhead}[1]{\par\noindent\textbf{#1}\ }

\title{
Beyond the Stability--Plasticity Frontier\\
in Streaming Target Speaker Extraction
}

\name{Yuesheng Ma \qquad Linyang He \qquad Nima Mesgarani}
\address{Department of Electrical Engineering, Columbia University, New York, NY, USA}

\begin{document}

\ninept
\maketitle
\fontsize{9.5}{11.6}\selectfont


\begin{abstract}
Streaming target speaker extraction must maintain a representation of whom to
extract while the target may fall silent, be masked by interference, or drift
acoustically away from enrollment. Existing systems typically hold this state
as a stored embedding updated by hand-designed rules. Across 22 configurations,
including confidence-gated and oracle-activity-gated updates, we show that
this family lies on a stability--plasticity frontier: even perfect
target-activity information cannot combine robustness to target absence with
adaptation to enrollment--mixture mismatch. We therefore meta-train
speaker-state dynamics through the closed streaming loop, exposing the updater
to its own contaminated evidence. Our proposed 41k-parameter anchored
fast-weights (AFW) memory moves beyond the measured heuristic frontier,
gaining 3.0\,dB over the best heuristic under severe mismatch while staying
within 0.9\,dB of static enrollment after 30\,s of absence, at under 5\%
runtime overhead. A gated recurrent unit (GRU) control confirms that the gain is not AFW-specific, while AFW is
smaller and more interpretable: under severe mismatch, its write residual
grows and aligns with the target rather than the interferer. Code is publicly available at \url{https://github.com/ym2976/anchor-fast-weight}.
\end{abstract}



\section{Introduction}

\begin{figure}[t]
\centering
\begin{tikzpicture}[
  x=1mm,y=.70mm,
  font=\fontsize{9}{10}\selectfont,
  axis/.style={-{Stealth[length=3.3pt,width=2.9pt]},draw=black!75,line width=.6pt},
  endpoint/.style={circle,fill=black!55,draw=white,line width=.5pt,inner sep=0pt,minimum size=1.8mm},
  labeltext/.style={align=center,inner sep=1pt}
]
\definecolor{conceptTeal}{HTML}{137F78}
\path[use as bounding box] (0,0) rectangle (86,38);

\fill[black!5] (15,7) -- (15,31.5) -- (21,31.5)
  .. controls (41,31.5) and (56,30) .. (66,11)
  -- (66,7) -- cycle;
\draw[axis] (15,7) -- (84,7);
\draw[axis] (15,7) -- (15,37);
\node[rotate=90,labeltext] at (8,22) {Retain identity};
\node[labeltext] at (51,2) {Adapt to mismatch};

\draw[black!58,line width=1.45pt] (21,31.5)
  .. controls (41,31.5) and (56,30) .. (66,11);
\node[endpoint] at (21,31.5) {};
\node[endpoint] at (66,11) {};
\node[labeltext,anchor=south,text=black!75] at (28,33) {Hold state};
\node[labeltext,anchor=west,text=black!75] at (67,13.2) {Update readily};
\node[labeltext,text=black!70] at (35,19.2) {Heuristic\\frontier};

\draw[-{Stealth[length=5pt,width=4.4pt]},conceptTeal,line width=1.6pt]
  (50,25) -- (70,30);
\fill[conceptTeal!12] (74,30.5) circle (3mm);
\node[text=conceptTeal,font=\fontsize{15}{15}\selectfont,inner sep=0pt]
  at (74,30.5) {$\star$};
\node[labeltext,anchor=south,text=conceptTeal,font=\fontsize{9}{10}\selectfont\bfseries]
  at (71,34.2) {Learned states};
\end{tikzpicture}
\caption{\textbf{Conceptual stability--plasticity frontier.}
Heuristic tuning trades retention for adaptation; learned state dynamics can
reach operating points beyond this boundary.}
\label{fig:concept}
\end{figure}
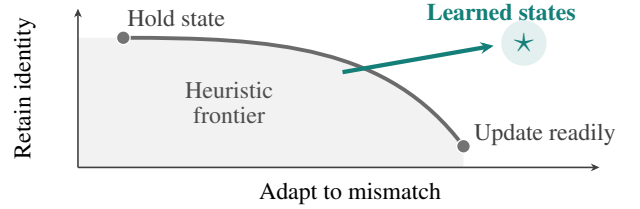

Target speaker extraction (TSE) uses an enrollment recording to specify whom
to extract from a mixture~\cite{zmolikova2023overview,
zmolikova2019speakerbeam,voicefilter,ge2020spex}. In streaming
systems~\cite{voicefilterlite}, that identity must remain useful beyond the
initial recording. Two situations demand opposite responses
(Fig.~\ref{fig:concept}). When the target sounds different from enrollment,
the state should adapt; when the target disappears, it should resist
interference. Keeping the cue fixed avoids state drift but forfeits adaptation,
whereas updating it from the separator's own imperfect output can reinforce a
wrong-speaker estimate. This creates a stability--plasticity
dilemma~\cite{grossberg}.

Recent work uses momentum~\cite{momuse}, memory banks~\cite{memo}, and
previously extracted speech~\cite{avtse_ar,evotse} to update speaker cues.
These systems establish the value of historical evidence, and autoregressive
or artifact-aware training can expose models to estimated
references~\cite{memo,avtse_ar,evotse}. However, state maintenance in these
systems is still largely governed by hand-designed momentum, gating, or
retrieval rules. It therefore remains unclear whether such update rules can
both preserve identity through target absence and adapt to
enrollment--mixture mismatch. This leads to our central question:
\emph{how can a speaker state adapt without losing identity?} The key
difficulty is that activity is not identity: even while the target is active,
the separator output can contain interferer evidence, so an activity gate may
still write the wrong speaker into state. Because this failure lives in the update rule, we freeze the
separation backbone and speaker encoder and vary only the state
mechanism. 

We make three contributions. First, we empirically characterize this trade-off
across 22 adaptive heuristic settings. The fact that oracle target-activity-gated updates form the heuristic frontier shows that even perfect timing does not determine whose evidence gets written. Second, we
show that meta-training state dynamics through the closed streaming loop can
move beyond this frontier by exposing the updater to the consequences of its
earlier decisions, consistent with training on self-generated
histories~\cite{dagger,schedsampling}. Third, we propose anchored fast weights (AFW), a
compact and interpretable realization of this principle. AFW performs online
adaptation through a content-addressed directional residual around a fixed
enrollment anchor.


\section{Anchored fast weights for closed-loop state learning}
\label{sec:method}

\begin{figure}[t]
\centering
\begin{tikzpicture}[
  x=1mm,y=1mm,font=\fontsize{9}{10}\selectfont,
  box/.style={draw,rounded corners=1.3pt,align=center,line width=.55pt,inner sep=3pt},
  frozen/.style={box,fill=black!6},
  memory/.style={box,line width=1.1pt},
  arr/.style={-{Stealth[length=3.8pt,width=3.2pt]},line width=.65pt},
  feedback/.style={-{Stealth[length=4.5pt,width=3.5pt]},line width=1.1pt},
  lab/.style={inner sep=1pt,fill=white}
]
\path[use as bounding box] (0,-.5) rectangle (86,49);
\node[frozen,minimum width=29mm,minimum height=10mm] (sep) at (27,35)
  {Separator $F$\\frozen: 2.9M};
\node[frozen,minimum width=19mm,minimum height=10mm] (enc) at (71,35)
  {Encoder $E$\\frozen: 0.26M};
\draw[arr] (0,35) -- node[lab,above] {$x_t$} (sep.west);
\draw[feedback] (sep.east) -- node[lab,above] {$\hat y_t$} (enc.west);
\node[memory,minimum width=29mm,minimum height=11mm] (mem) at (67,12)
  {AFW memory $\W_t$\\41k meta-params};
\draw[feedback] (enc.south) -- node[lab,right] {$\et$} (71,20) -- (mem.north);
\node[box,minimum width=29mm,minimum height=11mm] (read) at (27,12)
  {Anchor + residual\\$\norm(\anchor+\W_t\mathbf q)$};
\draw[feedback] (mem.west) -- (read.east);
\draw[feedback] (read.west) -- (5,12) -- (5,24) --
 node[lab,above,pos=.53] {$\st$ to chunk $t{+}1$} (27,24) -- (sep.south);
\node[anchor=south,align=center] at (43,42)
  {\textbf{Own output becomes future evidence}};
\node[anchor=north,align=center] at (43,3.5)
  {Train through 52--104 chunks; freeze $F$ and $E$};
\end{tikzpicture}
\caption{\textbf{Anchored fast weights (AFW) in the closed streaming loop.}
The output $\hat{\mathbf y}_t$ is re-encoded as evidence $\mathbf e_t$, which
updates the AFW memory for the next chunk. The state is read out as a residual
over the fixed enrollment anchor; $F$ and $E$ remain frozen.}
\label{fig:architecture}
\end{figure}
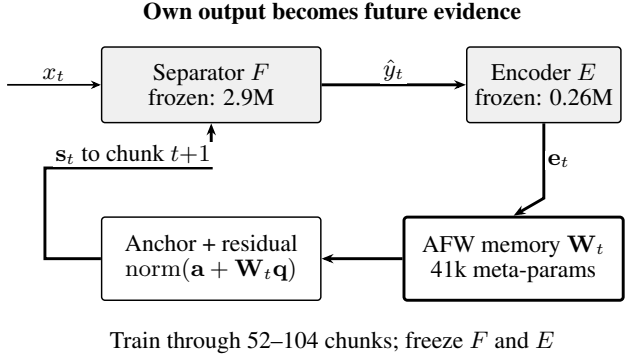

\pointhead{Closed-loop setting.}
Let $\anchor=E(\mathrm{enroll})$ denote the fixed identity anchor, $\st$ the
speaker state used for subsequent extraction, and
$\et=E(\hat{\mathbf y}_t)$ the evidence encoded from the current output.
Figure~\ref{fig:architecture} separates these roles. Importantly, the evidence
is not an independent observation: it comes from an output produced using the
previous state. A state update therefore changes the evidence available to
later updates, making state maintenance a closed-loop problem.

\pointhead{Frozen extraction modules.}
The frozen system comprises a separation backbone $F$ and speaker encoder $E$
(Fig.~\ref{fig:architecture}). The backbone encodes each 250\,ms mixture
chunk, estimates a target mask with a causal  Feature-wise Linear Modulation (FiLM)-conditioned Temporal
Convolutional Network (TCN)~\cite{convtasnet,film}, and decodes
$\hat{\mathbf y}_t$; $E$ embeds enrollment and $\hat{\mathbf y}_t$ in
$\mathbb R^{128}$. Both are pretrained with static enrollment and frozen
(2.9M/0.26M parameters), so all comparisons isolate the state mechanism.

\pointhead{Anchored fast weights (AFW).}
We use an associative memory~\cite{ba2016fastweights,schlag2021fwp} with
$\W_0=0$ and $\W_t\in\mathbb{R}^{128\times128}$. Evidence produces a key and
value, $\mathbf{k}_t=K\et$ and $\mathbf{v}_t=V\et$. A delta-rule write with
forgetting~\cite{gateddeltanet} updates the memory:
\begin{equation}
\W_t=(1-\lambda_t)\W_{t-1}
-\eta_t(\W_{t-1}\mathbf{k}_t-\mathbf{v}_t)\mathbf{k}_t^\top.
\label{eq:update}
\end{equation}
The write is a gradient step on
$\tfrac12\|\W\mathbf{k}_t-\mathbf{v}_t\|^2$, combined with matrix decay.
A 64-unit gate MLP takes $\mathbf e_t$,
$\cos(\mathbf e_t,\mathbf s_{t-1})$, and the clipped log-energy of
$\hat{\mathbf y}_t$, and outputs write and forgetting rates
$\eta_t\in(0,1)$ and $\lambda_t\in(0,0.1)$. A learned query $\mathbf q$ reads
the residual $\mathbf r_t=\W_t\mathbf q$ over the original enrollment:
\begin{equation}
\st=\norm\!\left(\anchor+\mathbf r_t\right).
\label{eq:readout}
\end{equation}
where $\norm(\mathbf z)=\mathbf z/\|\mathbf z\|_2$.
The anchor remains present in every readout, while the memory models a
context-dependent deviation from enrollment. The 41k trainable parameters are
$K,V,\mathbf q$, and the rate network. Following the fast-weights view of
test-time learning~\cite{ttt,titans}, these parameters learn \emph{how} to
update offline while $\W_t$ changes online. Inference requires only a rank-1
write and matrix--vector products, not autograd. Unlike model adaptation
through auxiliary speech-enhancement losses~\cite{sun2020ttt,tttse}, the
separator weights remain unchanged.

\pointhead{Closed-loop meta-training and GRU control.}
We backpropagate through 13--26\,s scenarios (52--104 chunks). Let
$\hat{\mathbf y}_{\mathcal A}$ and $\mathbf y_{\mathcal A}$ concatenate
target-active samples, and $\hat{\mathbf y}_{\bar{\mathcal A}}$ the
target-absent samples. We optimize
\begin{equation}
\mathcal L =
-\mathrm{SI\text{-}SNR}(\hat{\mathbf y}_{\mathcal A},
\mathbf y_{\mathcal A})
+0.1 P_{\mathrm{abs}},
\end{equation}
\begin{equation*}
P_{\mathrm{abs}} =
10\log_{10}\!\left(
\operatorname{mean}\hat{\mathbf y}_{\bar{\mathcal A}}^{2}+10^{-8}
\right).
\end{equation*}
No explicit speaker-identity loss is used. Training includes absences up to
4\,s and channel severities 0--3. To test whether the benefit of closed-loop
learning depends specifically on AFW, we train a 128-dimensional GRUCell with
the same objective and unroll, initialized from enrollment and read out as an
enrollment residual.


\section{Measuring the stability--plasticity frontier}
\label{sec:protocol}

\pointhead{Common protocol.}
We use LibriSpeech~\cite{librispeech} train-clean-360 for training and
speaker-disjoint test-clean for evaluation, with 150 sequences per condition.
The base condition has no scripted absence ($D{=}0$), no channel mismatch
($s{=}0$), and Target-to-Masker Ratio (TMR) $=0$\,dB. Each test changes one factor
at a time.

\pointhead{Two evaluation axes.}
\emph{Stability} measures SI-SNR improvement (SI-SNRi) after the target
returns: 5\,s of target-active speech, $D$\,s of absence, then 8\,s of
active speech, with $s{=}0$. We sweep
$D\in\{1,4,8,16,30,45,60\}$\,s; durations above 4\,s exceed the training
range. \emph{Plasticity} measures SI-SNRi under enrollment--mixture mismatch
with $D{=}0$. The in-mixture target is filtered with random spectral tilt and
peaking filters, scaled by $s\in\{1,2,3\}$; $s{=}3$ additionally applies a
telephone bandpass. Test filters use new parameter draws from the training
family.
The two scores assess separate conditions, not simultaneous mismatch and
absence. A configuration lies on the \emph{empirical heuristic frontier} if improving one
axis requires sacrificing the other. No monotonicity of an individual sweep is assumed.

\pointhead{State-update controls.}
Static conditioning keeps $\mathbf{s}_t=\mathbf{a}$. All adaptive heuristic
baselines use the same Exponential Moving Average (EMA) update,
$\mathbf{s}_t=\alpha\mathbf{s}_{t-1}+(1-\alpha)\mathbf{e}_t$, but differ in
when the update is applied. Plain EMA updates every chunk; confidence-gated
EMA updates only when
$\cos(\mathbf{e}_t,\mathbf{s}_{t-1})>\theta$; and oracle activity-gated EMA
updates only when the target is active, using ground-truth Voice Activity
Detection (VAD) labels to remove detection error~\cite{personalvad}. We
evaluate 9 plain-EMA, 8 confidence-gated, and 5 oracle-activity-gated settings
(22 adaptive heuristics), plus static enrollment. These are controlled state
rules on one backbone. An \emph{oracle-evidence} diagnostic feeds clean-target
encodings to the trained AFW state without changing the separator. It isolates
the headroom available from better evidence with the same trained state and
separator.

\pointhead{Metrics and reporting.}
SI-SNRi~\cite{leroux2019sdr} is evaluated on target-active speech; during
absence, we report output suppression in dB. Speaker confusion counts 1.6\,s
windows whose output is closer to the interferer than the target according to
an external ECAPA-TDNN~\cite{ecapa}, not the trained encoder. Learned main
results average three seeds. Ablation variants use one seed;
Table~\ref{tab:abl} separately reports sequence-level uncertainty for the full
system. For the AFW diagnostic, we compare $\mathbf r_t$ with target- and
interferer-offset directions relative to enrollment, random directions, and
its norm at $s=0$ and $s=3$. 
Additional tests vary TMR, training coverage, and 325 held-out measured
room impulse responses (RIRs) from RWCP~\cite{rwcp}, AIR~\cite{air}, and
REVERB~\cite{reverb}, with no test RIR seen during training.


\begin{table*}[t]
\centering
\caption{\textbf{One backbone, different speaker states.}
All separation scores are SI-SNRi in dB. Absence columns fix $s{=}0$;
mismatch columns fix $D{=}0$; TMR columns fix $D{=}0$ and $s{=}0$; TMR is
0\,dB except in its own columns. Suppression is measured during a 16\,s
absence; confusion is reported at $s{=}3$. Bold heuristic entries mark the
best marginal score on each main axis. Learned rows average three seeds.
Figure~\ref{fig:frontier} includes the full 22-setting sweep.}
\label{tab:main}

\setlength{\tabcolsep}{2.75pt}
\renewcommand{\arraystretch}{0.95}

\begin{tabular*}{\textwidth}
{@{\extracolsep{\fill}}lrrrrrrrrrrr@{}}
\toprule
& Base
& \multicolumn{3}{c}{Absence $D$ (s)}
& \multicolumn{3}{c}{Mismatch $s$}
& \multicolumn{2}{c}{TMR (dB)}
& Supp. & Conf.\\
\cmidrule(lr){3-5}
\cmidrule(lr){6-8}
\cmidrule(lr){9-10}
Speaker state
& & 4 & 16 & 30
& 1 & 2 & 3
& $-10$ & $-15$
& (dB) & (\%)\\
\midrule
Oracle VAD, $\alpha{=}.995$ & 9.9 & 9.0 & 8.3 & \textbf{7.2} & 9.6 & 9.1 & 3.8 & 13.4 & 13.2 & 8.0 & 28 \\
Static enrollment & 9.9 & 9.0 & 8.3 & 7.1 & 9.7 & 9.1 & 3.0 & 13.7 & 13.5 & 8.2 & 31 \\
Gated EMA, $\theta{=}.7$ & 9.9 & 9.0 & 8.2 & 7.0 & 9.6 & 9.1 & 3.0 & 13.7 & 13.5 & 8.2 & 31 \\
EMA, $\alpha{=}.99$ & 9.8 & 8.9 & 8.1 & 6.5 & 9.5 & 9.1 & 4.5 & 13.1 & 13.0 & 8.3 & 26 \\
Gated EMA, $\theta{=}.5$ & 9.3 & 8.4 & 7.6 & 5.3 & 9.1 & 8.7 & 3.4 & 12.7 & 12.5 & 8.1 & 30 \\
Oracle VAD, $\alpha{=}.90$ & 3.5 & 3.4 & 2.2 & 1.1 & 3.5 & 4.6 & \textbf{7.9} & 6.5 & 7.0 & 5.8 & 15 \\
Gated EMA, $\theta{=}.3$ & 5.5 & 4.6 & 0.5 & -2.2 & 5.8 & 5.8 & 6.7 & 8.1 & 8.5 & 7.2 & 19 \\
EMA, $\alpha{=}.90$ & 3.4 & 1.7 & -3.3 & -4.2 & 3.4 & 4.5 & 7.6 & 6.3 & 6.9 & 7.4 & 16 \\
EMA, $\alpha{=}.95$ & 6.1 & 5.0 & -2.1 & -5.9 & 6.1 & 6.6 & 7.7 & 9.1 & 9.4 & 8.6 & 16 \\
\midrule
\textbf{GRU state} & 9.7 & 9.0 & 8.5 & 7.3 & 9.6 & 9.5 & 10.5 & 13.4 & 13.7 & 18.7 & 5 \\
\textbf{AFW memory} & 9.8 & 8.8 & 7.8 & 6.2 & 9.6 & 9.6 & 10.9 & 13.6 & 14.0 & 17.2 & 4 \\
\addlinespace[2pt]
\quad + oracle evidence & 10.6 & 9.7 & 9.0 & 7.9 & 10.7 & 10.8 & 11.3 & 14.6 & 15.0 & 20.7 & 3 \\
\bottomrule

\end{tabular*}

\end{table*}


\section{Results}
\label{sec:results}

\begin{figure}[t]
\centering
\includegraphics[width=.95\columnwidth]{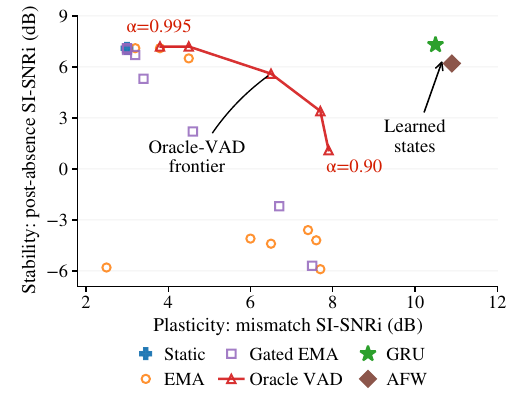}
\caption{\textbf{The measured frontier.}
Oracle-activity gating defines the empirical heuristic frontier; AFW and the
GRU control lie beyond it.}
\label{fig:frontier}
\end{figure}

\pointhead{Oracle timing defines the heuristic frontier.}
The frontier curve in Fig.~\ref{fig:frontier} is traced entirely by oracle
target-activity-gated heuristics. Table~\ref{tab:main} shows the corresponding
values: the
aggressive oracle gate reaches 7.9\,dB under severe mismatch but only
1.1\,dB after 30\,s absence, whereas the conservative setting reaches
7.2\,dB after absence but only 3.8\,dB under mismatch. Confidence gating
offers no escape: $\theta=.7$ is static-like, while $\theta=.3$ raises
mismatch performance to 6.7\,dB but lowers post-absence performance to
$-2.2$\,dB. Without scripted absence, the oracle VAD gate opens on 98.2\% of
chunks and matches EMA at $\alpha=.90$ within 0.1\,dB (3.5 versus 3.4).
Perfect update timing therefore improves the heuristic trade-off but does not
resolve whose evidence is written.

\pointhead{AFW moves beyond the frontier; GRU verifies the learning principle.}
AFW reaches 10.9\,dB under severe mismatch and 6.2\,dB after 30\,s
absence, compared with the best heuristic score on each axis, 7.9 and
7.2\,dB, achieved by different configurations.
The GRU control reaches 10.5 and 7.3\,dB on the same axes, confirming that
moving beyond the heuristic frontier is not specific to fast weights. AFW
therefore favors adaptation with 41k state parameters, whereas the
115k-parameter GRU favors retention. This improvement is consistent across
secondary metrics: both learned states remain robust at TMR $=-10/-15$\,dB,
reduce speaker confusion to 4--5\%, and increase absence suppression to
17.2--18.7\,dB.

\begin{figure}[t]
\centering
\includegraphics[width=\columnwidth]{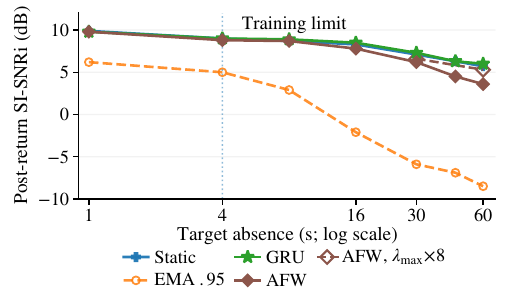}
\caption{\textbf{Retention extrapolates beyond the training range.}
Post-return SI-SNRi versus absence length; training saw only $D\leq4$\,s.
GRU stays close to static through 60\,s. The AFW
$\lambda_{\max}\times8$ points show an inference-time retention dial at
$D=30$ and $60$\,s.}
\label{fig:absence}
\end{figure}

\pointhead{Retention extends beyond the training duration.}
Figure~\ref{fig:absence} isolates the stability axis. Although training
contains absences only up to 4\,s, the GRU architecture control remains close
to static enrollment through 60\,s (6.0 versus 5.7\,dB), whereas AFW reaches
3.6\,dB. AFW additionally exposes an inference-time retention control:
raising its forgetting-rate cap shifts performance from 6.2 to 6.6\,dB at
30\,s and from 3.6 to 5.3\,dB at 60\,s without retraining, for only a
0.2\,dB cost at $s=3$.

\begin{figure}[t]
\centering
\includegraphics[width=\columnwidth]{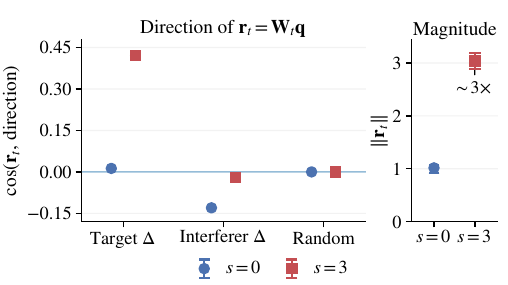}
\caption{\textbf{What AFW writes.}
For the residual $\mathbf r_t=\W_t\mathbf q$ added to the enrollment anchor,
the left axis shows directional cosine and the right axis its norm. Under
severe mismatch, AFW writes a larger correction aligned with the target
offset rather than the interferer.}
\label{fig:mechanism}
\end{figure}

\pointhead{AFW reveals how the frontier is crossed.}
AFW makes the learned update directly inspectable through the residual
$\mathbf r_t=\W_t\mathbf q$ added to the enrollment anchor. If the memory
merely wrote more of whatever evidence arrived, this residual would not
preferentially point toward the target. Figure~\ref{fig:mechanism} shows the
opposite. In matched speech ($s=0$), the residual is small
($\|\W_t\mathbf q\|=1.016\pm0.282$), has almost no target-offset alignment
($0.013\pm0.007$), negative interferer alignment ($-0.128$), and
random-direction cosine near zero ($-0.006$). Under severe mismatch ($s=3$),
the residual grows substantially ($3.100\pm0.637$) and aligns strongly with
the target offset ($0.419\pm0.003$), while remaining near orthogonal to the
interferer ($-0.018$) and random directions ($-0.008$). AFW therefore applies
a content-addressed directional correction rather than simply increasing the
amount written. The fixed-rate ablation further separates direction from
scalar scheduling: replacing the gate network with constants
($\eta=.10$, $\lambda=.010$) changes mismatch and post-absence SI-SNRi by
only $-0.2$\,dB each. The learned keys, values, query, and prediction error
still determine the correction direction, whereas EMA-style heuristics can
only control how much evidence enters the state.

\begin{table}[t]
\centering
\caption{\textbf{AFW ablations} under the same conditions as
Table~\ref{tab:main}: mismatch and confusion at $s{=}3$, absence at
$D{=}30$, and suppression at $D{=}16$. The full system averages three
seeds; variants are single-seed.}
\label{tab:abl}

\setlength{\tabcolsep}{2.75pt}
\renewcommand{\arraystretch}{0.95}

\begin{tabular*}{\columnwidth}
{@{\extracolsep{\fill}}lrrrr@{}}
\toprule
AFW variant & $s{=}3$ & $D{=}30$ & Supp. & Conf.\\
& (dB) & (dB) & (dB) & (\%)\\
\midrule
Full system & 10.9 & 6.2 & 17.2 & 4 \\
Fixed scalar rates & 10.7 & 6.0 & 17.4 & 5 \\
No enrollment anchor & 10.8 & 5.0 & 17.1 & 4 \\
No suppression loss & 11.0 & 5.7 & 7.8 & 5 \\
No channel augmentation & 4.7 & 7.3 & 17.3 & 23 \\
Channel aug.\ $s\leq2$ & 6.3 & 7.2 & 16.9 & 18 \\
Oracle evidence & 11.3 & 7.9 & 20.7 & 3 \\
\bottomrule

\end{tabular*}

\end{table}

\begin{figure}[t]
\centering
\includegraphics[width=\columnwidth]{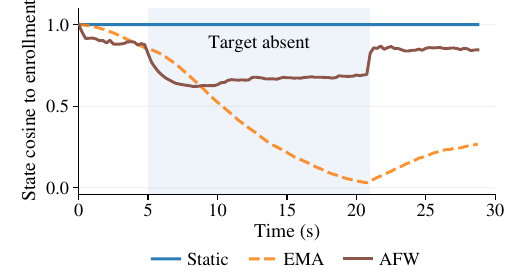}
\caption{\textbf{A target can return before its identity state recovers.}
A representative 16\,s absence drives the EMA state away from enrollment,
and it remains displaced after the target returns. AFW retains more of the
original state and moves back toward the anchor.}
\label{fig:trace}
\end{figure}

\pointhead{The anchor and suppression objective have distinct roles.}
Figure~\ref{fig:trace} illustrates the identity-recovery problem over time:
EMA's state cosine approaches 0.03 during absence and remains displaced after
the target returns, whereas AFW retains more of the original state and moves
back toward the anchor. The ablations in Table~\ref{tab:abl} separate the
roles of the anchor and suppression objective. Removing the enrollment anchor
reduces post-absence SI-SNRi from 6.2 to 5.0\,dB while mismatch and
suppression change little (10.9 to 10.8\,dB and 17.2 to 17.1\,dB), showing
that the residual can still adapt but the fixed identity reference supports
recovery. Conversely, removing the absence-power loss lowers suppression from
17.2 to 7.8\,dB, with smaller changes in mismatch and post-absence SI-SNRi
(to 11.0 and 5.7\,dB). The anchor therefore chiefly supports post-absence
identity recovery, whereas the suppression objective chiefly controls output
during target absence.

\pointhead{Training coverage controls AFW's operating point.}
At $s=3$, performance follows
$3.04\rightarrow4.68\rightarrow6.29\rightarrow10.87$\,dB for static
conditioning, AFW trained without channel augmentation, AFW trained only
through $s\leq2$, and AFW trained through $s\leq3$, respectively. The
$s\leq2$ model is evaluated at the unseen severity $s=3$ and still exceeds
static by 3.25\,dB, demonstrating extrapolation beyond the trained severity
range. Broader training coverage then moves AFW to a stronger operating point,
an axis unavailable to fixed heuristics because additional training diversity
cannot alter a hand-specified update rule.

The same pattern extends across corruption families. On measured room
responses absent from training, base AFW scores 3.3\,dB
(static: 3.3; EMA: 5.8; oracle evidence: 7.9). Expanding training to two and
three corruption families raises AFW to 3.7 and 4.0\,dB, and adding synthetic
reverberation~\cite{ko2017rir} reaches 4.6\,dB.
Separately, clean-target evidence raises the 30\,s score from 6.2 to
7.9\,dB, leaving evidence quality as additional headroom.

\pointhead{AFW remains a lightweight state mechanism.}
AFW adds 41k trainable parameters. Its measured real-time factor is 0.074 on
an A100 and 0.235 on CPU, versus 0.071 and 0.241 for static conditioning,
corresponding to under 5\% measured runtime difference. On matched,
short-form Libri2Mix~\cite{librimix} test-clean, AFW scores 7.93\,dB versus
8.40\,dB for static conditioning. This localizes the contribution: AFW is not
intended as a replacement separator, but as a lightweight state mechanism for
conditions that require adaptation across enrollment mismatch and prolonged
target absence.


\section{Conclusion}

We identify a measurable stability--plasticity frontier in streaming
speaker-state maintenance whose empirical heuristic boundary is traced by
oracle target-activity-gated updates. Closed-loop meta-training can move
speaker-state dynamics beyond this boundary; a GRU architecture control shows
that this principle is not specific to one state representation. Our proposed
AFW realizes the same principle with a compact, interpretable state: a fixed
enrollment anchor is corrected online by a content-addressed directional
residual that aligns with the target offset under mismatch. Its ablations
separate directional adaptation, identity retention, and output suppression,
while the training-coverage experiments show that learned state dynamics gain
an additional design axis unavailable to fixed update rules. AFW therefore
provides a lightweight and inspectable realization of state adaptation beyond
the measured heuristic frontier.
\clearpage
\section{ACKNOWLEDGEMENTS}
The conceptual illustration in Fig.~\ref{fig:concept} and the architecture diagram in Fig.~\ref{fig:architecture} were generated with ChatGPT (OpenAI), which was also used to refine the manuscript's language. The manuscript was written by the authors, who reviewed the final text and figures and take full responsibility for their content.

This work was funded by the National Institutes of Health (NIH-NIDCD) and a grant from Marie-Josee and Henry R. Kravis.

\fontsize{9}{10.5}\selectfont


\begin{thebibliography}{99}
\setlength{\emergencystretch}{2em}
\bibitem{zmolikova2023overview} K.~\v{Z}mol\'{i}kov\'{a}, M.~Delcroix, T.~Ochiai, K.~Kinoshita, J.~\v{C}ernock\'{y}, and D.~Yu, ``Neural target speech extraction: An overview,'' \emph{IEEE Signal Process.\ Mag.}, vol.~40, no.~3, 2023.
\bibitem{zmolikova2019speakerbeam} K.~\v{Z}mol\'{i}kov\'{a}, M.~Delcroix, K.~Kinoshita, T.~Ochiai, T.~Nakatani, L.~Burget, and J.~\v{C}ernock\'{y}, ``SpeakerBeam: Speaker aware neural network for target speaker extraction in speech mixtures,'' \emph{IEEE J.\ Sel.\ Topics Signal Process.}, vol.~13, no.~4, 2019.
\bibitem{voicefilter} Q.~Wang et al., ``VoiceFilter: Targeted voice separation by speaker-conditioned spectrogram masking,'' in \emph{Interspeech}, 2019.
\bibitem{ge2020spex} M.~Ge, C.~Xu, L.~Wang, E.~S.~Chng, J.~Dang, and H.~Li, ``SpEx+: A complete time domain speaker extraction network,'' in \emph{Interspeech}, 2020.
\bibitem{voicefilterlite} Q.~Wang et al., ``VoiceFilter-Lite: Streaming targeted voice separation for on-device speech recognition,'' in \emph{Interspeech}, 2020.
\bibitem{momuse} J.~Li, K.~Zhang, S.~Wang, K.~A.~Lee, M.-W.~Mak, and H.~Li, ``MoMuSE: Momentum multi-modal target speaker extraction for real-time scenarios with impaired visual cues,'' arXiv:2412.08247, 2024.
\bibitem{memo} J.~Li, W.~Wu, S.~Wang, Z.~Pan, K.~A.~Lee, H.~Meng, and H.~Li, ``MeMo: Attentional momentum for real-time audio-visual target speaker extraction under impaired visual conditions,'' arXiv:2507.15294, 2025.
\bibitem{avtse_ar} Z.~Pan et al., ``Online audio-visual autoregressive speaker extraction,'' in \emph{Interspeech}, 2025.
\bibitem{evotse} Z.~Liu, Z.~Wang, X.~Li, Y.~Zhu, S.~Wang, L.~Xiao, and L.~Xie, ``EvoTSE: Evolving enrollment for target speaker extraction,'' arXiv:2604.06810, 2026.
\bibitem{grossberg} S.~Grossberg, ``Competitive learning: From interactive activation to adaptive resonance,'' \emph{Cognitive Science}, vol.~11, no.~1, 1987.
\bibitem{dagger} S.~Ross, G.~J.~Gordon, and J.~A.~Bagnell, ``A reduction of imitation learning and structured prediction to no-regret online learning,'' in \emph{AISTATS}, 2011.
\bibitem{schedsampling} S.~Bengio, O.~Vinyals, N.~Jaitly, and N.~Shazeer, ``Scheduled sampling for sequence prediction with recurrent neural networks,'' in \emph{NeurIPS}, 2015.
\bibitem{convtasnet} Y.~Luo and N.~Mesgarani, ``Conv-TasNet: Surpassing ideal time--frequency magnitude masking for speech separation,'' \emph{IEEE/ACM Trans.\ Audio, Speech, Lang.\ Process.}, vol.~27, no.~8, 2019.
\bibitem{film} E.~Perez, F.~Strub, H.~de~Vries, V.~Dumoulin, and A.~Courville, ``FiLM: Visual reasoning with a general conditioning layer,'' in \emph{AAAI}, 2018.
\bibitem{ba2016fastweights} J.~Ba, G.~E.~Hinton, V.~Mnih, J.~Z.~Leibo, and C.~Ionescu, ``Using fast weights to attend to the recent past,'' in \emph{NeurIPS}, 2016.
\bibitem{schlag2021fwp} I.~Schlag, K.~Irie, and J.~Schmidhuber, ``Linear transformers are secretly fast weight programmers,'' in \emph{ICML}, 2021.
\bibitem{gateddeltanet} S.~Yang, J.~Kautz, and A.~Hatamizadeh, ``Gated delta networks: Improving Mamba2 with delta rule,'' in \emph{ICLR}, 2025.
\bibitem{ttt} Y.~Sun et al., ``Learning to (learn at test time): RNNs with expressive hidden states,'' in \emph{ICML}, 2025.
\bibitem{titans} A.~Behrouz, P.~Zhong, and V.~Mirrokni, ``Titans: Learning to memorize at test time,'' arXiv:2501.00663, 2025.
\bibitem{sun2020ttt} Y.~Sun, X.~Wang, Z.~Liu, J.~Miller, A.~A.~Efros, and M.~Hardt, ``Test-time training with self-super\-vision for generalization\linebreak under distribution shifts,'' in \emph{ICML}, 2020.
\bibitem{tttse} A.~Behera, R.~A.~Easow, V.~Parvathala, and K.~S.~R.~Murty, ``Test-time training for speech enhancement,'' in \emph{Interspeech}, 2025.
\bibitem{librispeech} V.~Panayotov, G.~Chen, D.~Povey, and S.~Khudanpur, ``LibriSpeech: An ASR corpus based on public domain audio books,'' in \emph{ICASSP}, 2015.
\bibitem{personalvad} S.~Ding, Q.~Wang, S.-Y.~Chang, L.~Wan, and I.~Lopez Moreno, ``Personal VAD: Speaker-conditioned voice activity detection,'' in \emph{Odyssey}, 2020.

\bibitem{leroux2019sdr} J.~Le~Roux, S.~Wisdom, H.~Erdogan, and J.~R.~Hershey, ``SDR---half-baked or well done?'' in \emph{ICASSP}, 2019.

\bibitem{ecapa} B.~Desplanques, J.~Thienpondt, and K.~Demuynck, ``ECAPA-TDNN: Emphasized channel attention, propagation and aggregation in TDNN based speaker verification,'' in \emph{Interspeech}, 2020.
\bibitem{rwcp}
S.~Nakamura, K.~Hiyane, F.~Asano, T.~Nishiura, and T.~Yamada,
``Acoustical sound database in real environments for sound scene understanding and hands-free speech recognition,''
in \emph{Proc. LREC}, 2000, pp. 965--968.

\bibitem{air}
M.~Jeub, M.~Sch{\"a}fer, and P.~Vary,
``A binaural room impulse response database for the evaluation of dereverberation algorithms,''
in \emph{Proc. IEEE Int. Conf. Digital Signal Processing (DSP)}, 2009, pp. 1--4.

\bibitem{reverb}
K.~Kinoshita, M.~Delcroix, T.~Yoshioka, T.~Nakatani, E.~A.~P.~Habets,
R.~Haeb-Umbach, V.~Leutnant, A.~Sehr, W.~Kellermann, R.~Maas,
S.~Gannot, and B.~Raj,
``The REVERB challenge: A common evaluation framework for dereverberation and recognition of reverberant speech,''
in \emph{Proc. IEEE Workshop Applications of Signal Processing to Audio and Acoustics (WASPAA)}, 2013, pp. 1--4.
\bibitem{ko2017rir} T.~Ko, V.~Peddinti, D.~Povey, M.~L.~Seltzer, and S.~Khudanpur, ``A study on data augmentation of reverberant speech for robust speech recognition,'' in \emph{ICASSP}, 2017.
\bibitem{librimix} J.~Cosentino, M.~Pariente, S.~Cornell, A.~Deleforge, and E.~Vincent, ``LibriMix: An open-source dataset for generalizable speech separation,'' arXiv:2005.11262, 2020.

\end{thebibliography}
\end{document}